# Text Finder Application for Android

Dr. Milind Godase[1], Dr. Chandrani Singh[2], Kunal Dhongadi[3]

Email: [milindgodase@sinhgad.edu](milindgodase@sinhgad.edu)[1], [directormca_siom@sinhgad.edu](directormca_siom@sinhgad.edu)[2],

[Kunaldhongadi@gmail.com](Kunaldhongadi@gmail.com)[3]

**Sinhgad Institute of Management, Vadgaon (Bk), Pune (India)**

**Abstract**

A Text Finder, an android application that utilizes Optical Character Recognition (OCR) technology with the help of Google's Cloud Vision API to extract text from images taken with the device camera or from existing images in the user's phone. The extracted text can be saved to the device storage where all previous extracts can be easily accessed on a user-friendly interface. The application also features editing, deletion and sharing options for the extracted text. The user interface is user-friendly, making the application accessible to students, professional and organizations for a variety of purposes, including document scanning, data entry, and information retrieval. Manual extraction of text by typing or writing from images can be very time-consuming and can be prone to errors. This application is an efficient and simple solution for extracted texts and organizing important information from the photos. This paper describes the technical details of the OCR technology and Google's ML Kit Text Recognition API used in the application, as well as the design, implementation and evaluation of the application in terms of performance and accuracy. The research also explores the key objectives and benefits of Text Finder, such as reducing the time and effort required and increasing the efficiency of document-based tasks.

## 1. INTRODUCTION

Text recognition, commonly known as Optical Character Recognition (OCR), is a technology that recognizes text from images and converts the recognized text into a machine readable text [1]. This technology has been widely adopted in various applications such as document scanning, data entry, and information retrieval. OCR technology works by analysing the image and identifying patterns and features that correspond to characters, numbers, and symbols. These patterns are then compared against a pre-defined set of templates, which are used to determine the character or symbol represented by each pattern. In general, recognizing texts consists of various steps such as preparing the image for analysis, splitting the image into smaller segments containing individual characters, extracting characteristics from the segments that correspond to characters, comparing the extracted features against pre-defined templates, training the algorithm using labeled image data, testing the algorithm using a separate dataset, and making any necessary adjustments. Recognition algorithm can be different based on the type of image, font, and text, and it can be fine-tuned with a specific dataset that contains images similar to the ones that the OCR will be applied on.

As the amount of information stored in digital format continues to grow, the need for efficient and accurate text extraction and recognition has increased significantly. Traditional methods of data entry, such as manual typing, are time-consuming and prone to errors. OCR technology has emerged as a solution to this problem, by automating the process of text extraction and converting it into machine-readable format [2].

With the help of Machine Learning, the field of OCR is continuously advancing, which is making the OCR more accurate and efficient. One example of this is ML Kit by Google, which is a mobile SDK that includes text recognition APIs. This SDK allows developers to integrate OCR technology into mobile applications, which can be used for a variety of tasks such as business card scanning, document scanning, and text extraction from images. The text recognition API from ML kit segments texts into blocks, lines, elements and symbols.

Text Finder is an Android app that makes use of such OCR technology from Google's ML Kit making it easy to capture and organize important information from any photo. Text Finder quickly and accurately extracts text from any photo, making it simple to save, view, and share with others. The application's user-friendly interface allows users to easily manage all of their previously recognized text, so they can find what they need, when they need it. The app also offers seamless sharing capabilities, allowing users to share their recognized text with others via email, text, or social media. Without the right tools, users must manually transcribe text from photos, which can be error-prone and difficult to share with others. This makes it difficult for users to stay organized and stay on top of their work, and can lead to lost or forgotten information. Text Finder is therefore a very valuable tool for anyone looking to capture and organize important information from any photo. The app's intuitive, minimalistic and simple user-friendly interface allows users of all kinds to easily extract text from image, manage all of their recognized text, and share it through multiple options.

## 2. LITERATURE REVIEW

With rising use of digital devices like digital cameras, mobile phones, PDAs, content based image analysis methods have caught serious concentration in the current past years. An image text information mining system is divided into four stages: text detection, text identification, text localization, text mining as well as text enhancement. Among these stages, text detection and text localization are significant. Numerous methods are projected for text detection with text localization problems [3], a number of them achieved high-quality results for specific relevance. The presented methods of text detection with text localization can be generally categorized into two groups: Section based plus connected component (CC) based text detection. Various issues in addition to challenges specifically for Urdu text based on OCR have been discussed by Zaki et al. [4]. Sharma et al. [5], works on OCR procedure based on Convolutional Neural Network (CNN), their method uses this OCR method to mine information of a student filled in a particular form. This form contains 175 cells. A quantity of these cells are filled with capital alphabets, others are few numbers. It also explains how to identify each cell using the feature and CNN. Their method has achieved a good accuracy for both numerical as well as alphabetical data.

Prakash et al. [6], discussed with the OCR challenge for the Telugu, they have created a database for Telugu, a deep learning Technique, in addition to a consumer server solution for the online procedure of the algorithm. Moteelal and Murthy [7] have proposed a framework of continuous content locations that usually deal with different ideas and recognize content. This technique depends on the relevant section. This approach can deal with the content on dispose image plus not bright districts. They build up windows based applications that can track content as well as isolate content in this field. They use Windows telephone of 8.1 rendition, OCR, C#. This approach is a direct and open source. Mol et al. [8], provides a method for text detection as well as recognition in pictures. This method emphasizes Fractional Poisson enhancement for removing Laplacian noise of the input image. Next, the Maximally Stable Extremal Regions (MSER) emphasized on the edge of the preprocessed image. Local filtering is used to filter areas of non-texture in addition to be recognized by an OCR. The results of this method are better than previous methods in terms of Peak Signal to Noise Ratio (PSNR) and Structural Similarity (SSIM) calculations. This method has used standard ICDAR dataset that have nearly real time images. Purkaystha et al. [9], used Convolutional Neural Network (CNN) for Bengali handwritten character detection with satisfactory accuracy. Razik et al. [10], equipped a database of handwritten Bangla numeral named SUST:BNHD. He used deep learning particularly CNN for recognized work. Besides, various researchers have developed different mobile apps for the purpose of text detection. In recent years, the investigation took into account the words and writings from the camera of cell phones has been organized to translate those words and texts into different languages. One such mobile based approach to convert Thai script into a Malay script has been presented by Aini et al. [11], works efficiently in offline mode. This system consists of three stages of translation, dictionary development, and display result. Yu and Wan [12] uses smart phones to detect Chinese text. The system sets 8 different sizes on 7 common Fonts and 3755 characters. This system generally claims to detect Chinese texts from natural scene. Velumurugan et al. [13], suggest a smart reader based on the problem using Raspberry Pie and use the simulation of

OCR using Matlab. Captured images are sent for preprocessing, where functions such as noise removal plus skew correction are performed.

The image is light weight in addition to binarization is done. After this, image is approved in a segmentation phase, where the image breaks into characters. Verma et al. [14], introduced an OCR system to change photos in text. Synthetic analysis consists of three stages: Extraction of character, Recognition as well as post processing. In the recognition stage, the template largest correlation is referred to as a character in the image. Rao et al. [15], suggest a Neural Network (NN) technology for the handwritten OCR. Tamil text image are scanned, filtered and pre-processed. Preprocessing consists of character segmentation and picture binarization. The features are extracted and the picture is sent for post processing which converts the picture into text. Text is then fed to a speech engine which generates synthesize Tamil speech as output consists of a multipurpose response via pulsation motors, a new twin material case design in addition to a high resolution mini video cam. Kumari et al. [16], investigated the methods and strategies in achieving the role, in an efficient tracking algorithm, the traditional approaches divide the picture into associate pictures, which are then freely rated. The second class of segmentation parts clearly illustrated by the collected spatial features. The third class is between the primary two and is used with recombination. Farhat et al. [17], introduced a mixed feature extraction method. This article describes a set of zoning, vector crossing techniques. Zoning techniques divide the whole picture into a small picture. Then data attributes (for instance, role density) are calculated for each zone. Vector crossing technique is used to extract the features. The thought of this method is to calculate the number that crosses character pixels in the vector (horizontal and vertical) image. With crossing vector technology, it can be helpful with further technologies. Characters can't be recognized as a vector crossing alone. This method is also useful for other ways. Because it only collects some data features. The cons of these technique are explained by pleasing the arrangement of these two techniques. Rownak et al. [18], proposed a study for efficient segmentation of Bangali characters on paper documents by means of Curvy Scan. Ayyaz et al. [19], planned a hybrid feature mining process, multi-class SVM is used for recognition works, good accuracy for numbers and English characters has been achieved in their study. Singh et al. [20], used SVM to recognize the optical character, they removed the character restrictions using Moore Neighbor's tracing and then applied the chin code.

3. **RESEARCH OBJECTIVES**

1. To present a detailed analysis of the Text Finder application and its capabilities, including its use of OCR technology by Cloud Vision API for text extraction.

2. To evaluate the performance and accuracy of the Text Finder application in extracting text from images.

3. To explore the potential uses and benefits of the Text Finder application for individuals and organizations, such as reducing the time and effort required for manual data entry and increasing the efficiency of document-based tasks.

4. To understand the working of the application

5. To discuss recommendations for future improvements and developments of the Text Finder application.

6. To evaluate the implementation and accuracy of the Google's Cloud Vision API used in the Text Finder application.

4. SCOPE

The scope of this research is to study the Text Finder application, which is an android application that allows extracting text and organizing important information/ texts from the image. The research will focus on evaluating the performance, accuracy, usability and potential uses of the Text Finder application. It will provide a comprehensive analysis of the Text Finder application, including its use of OCR technology from Cloud Vision's API for text extraction, its performance and accuracy, and its potential uses and benefits for individuals and organizations. The research will also focus on evaluating the user-experience of the Text Finder application and its implementation of the API, which helps in analyzing the image, extracting the text and recognizing the characters.

5. RESEARCH METHODOLOGY

1. **Application Analysis -** Text Finder uses Text Recognition API from Google's Cloud Vision Machine Learning kit for extracting text from a given image through internet. The interface provides features to save, organize, modify, delete and share extracted texts.

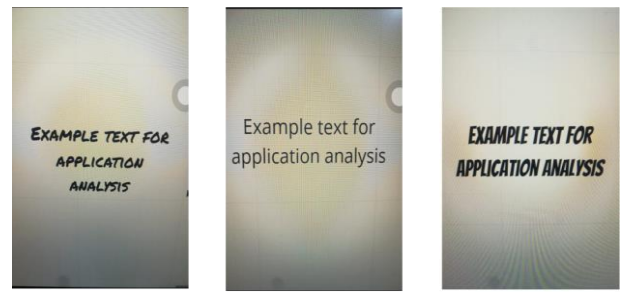

Image 1.1  Image 1.2  Image 1.3

| Image | Expected Output | Output |
|---|---|---|
| 1.1 | EXAMPLE TEXT FOR APPLICATION ANALYSIS | EXAMPLE TEXT FOR APPLICATION ANALYsIs |
| 1.2 | Example text for application analysis | Example text for application analysis |
| 1.3 | EXAMPLE TEXT FOR APPLICATION ANALYSIS | EXAMPLE TEXT FOR APPLICATION ANALVSIS |

Table 1: Testing of Images – 1

The above table consists of results from testing the application on various types of images with different fonts. The result displays a bit inaccurate output for Image 1.1 and Image 1.3.

2. **Performance Evaluation -** Results from testing the application on various types of images with different font styles (3D, curved and texts from different angles).

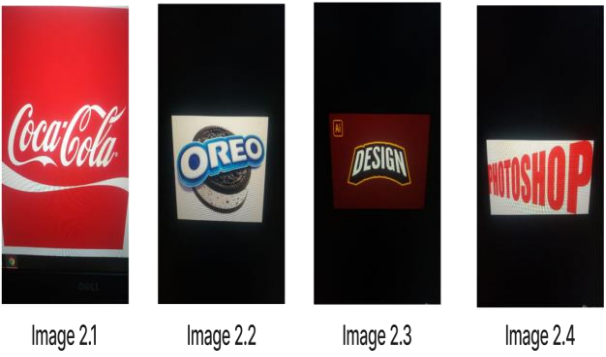

Image 2.1    Image 2.2    Image 2.3    Image 2.4

| Image | Expected Output | Output |
|---|---|---|
| 2.1 | CocaCola | DGLL |
| 2.2 | OREO | REO |
| 2.3 | Ai DESIGN | Ai DESIGN |
| 2.4 | Photoshop | IOSHO |

Table 2: Testing of Images – 2

The above table displays inaccurate outputs for all of the images except for Image 2.3.

3. **Language Text -** The Cloud Vision API is capable to automatically detect languages from the given image. Below are the results from images consisting of different languages.

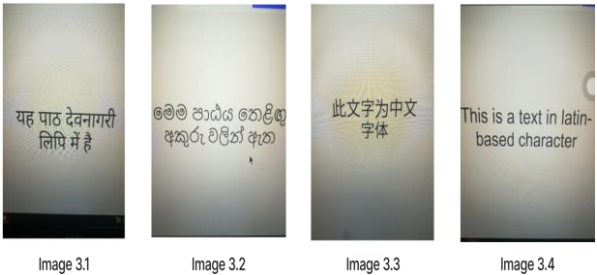

Image 3.1    Image 3.2    Image 3.3    Image 3.4

| Image | Expected Output | Output |
|---|---|---|
| 3.1 | यह पाठ देवनागरी लिपि में है | यह पाठ देवनागरी लिपि में है |
| 3.2 | මෙම පාඨය තෙළිඟු අකුරු වලින් ඇත | 18c |
| 3.3 | 此文字为中文字体 | [Returned null] |
| 3.4 | This is a text in Latin-based character | This is a text in Latin-based character |

Table 3: Testing of Images – 3

The above table evaluates that only Latin-based and Devanagari characters are extracted from the image

4. **Effect of image quality -** The image and its quality is one of the very important factor in text recognition. Below are the results extracted from images with varying quality images such as low resolution, blurry, low light and images clicked from different angles.

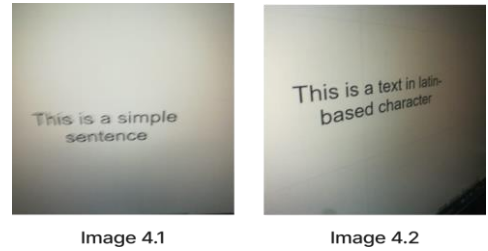

Image 4.1    Image 4.2

| Image | Image Description | Expected Output | Output |
|---|---|---|---|
| 4.1 | Blurry, Low res | This is a simple sentence | This is a simple Sentence |
| 4.2 | Different Angle | This is a text in Latin - based character | This is a text in Latin-based character |

Table 4: Testing of Images – 4

## 6. FLOWCHART DIAGRAM

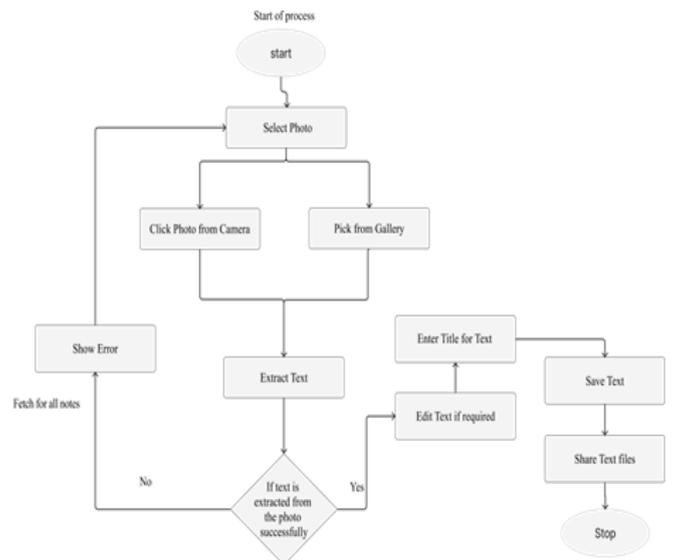

Figure – 1: Process - Flowchart

## 7. LIMITATIONS

1. No options to backup all notes at once.
2. Extracted text is not accurate and requires user to check and correct if necessary.
3. No option to extract particular text in the image or crop the image as per required.
4. Only English and Devanagari fonts can be detected from the image.

## 8. FUTURE ENHANCEMENTS

1. **Photo cropping options**- To allow users to crop the image according to the text they want to extract. This will add more flexibility and save time.

2. **Backup Options/ Store data in cloud** - To allow users to backup all their notes on their google drive account. Similarly, users could restore all their notes from their google drive account when they install the app on a new phone.

3. **Folder system**- To allow users to create folders and save texts. For e.g. All the texts related to college such as assignments and notes can be stored in the college folder. This will make organizing data easy.

## 9. CONCLUSION

In conclusion, the Text Finder app has been found to be a useful tool for extracting text from images. The results of the performance evaluation showed that the app has a high accuracy and precision but can struggle based on certain font types, text styles and the image quality. Additionally, the study found that the Text Finder app is only capable of extracting Latin-based characters and Devanagari fonts. While the features of this application has a wide range of potential uses and benefits for both individuals and organizations and it can save time and effort required for manual data entry and increase the efficiency of document-based tasks, the accuracy of the text recognition API from Google Cloud Vision heavily relies on the quality of the image and the format of text in the image. For these reasons, the outcome cannot be 100% accurate for every cases.

The research also found that the OCR technology by Google's Cloud Vision used in the Text Finder app is efficient and accurate in most cases. The application allows user to modify or correct any mistakes from the extracted texts. The study also identified some areas for improvement, such as expanding the languagee support, improving accuracy for variety of font styles and improving the text extraction from poor quality images. Overall, the research demonstrates the effectiveness and potential of the Text Finder app, and suggests future developments and improvements to enhance its capabilities.